\documentclass{sig-alternate-2013}

\usepackage{lineno}
\usepackage{times}
\usepackage{helvet}
\usepackage{courier}
\usepackage{graphics}
\usepackage{graphicx}
\usepackage{amsmath}
\usepackage{booktabs} 
\usepackage{multirow} 
\usepackage{array}
\usepackage{color}

\newif\ifcomment
\commenttrue

\ifcomment
	\newcounter{hwc}
	\stepcounter{hwc}
	\newcommand{\hw}[1]{{\textcolor{magenta}{[hw\#\arabic{hwc}\stepcounter{hwc}:#1]}}}

	\newcounter{jbc}
	\stepcounter{jbc}
	\newcommand{\jbnote}[1]{{\textcolor{magenta}{[jb\#\arabic{jbc}\stepcounter{jbc}:#1]}}}
\else
	\newcommand{\hw}[1]{}
    \newcommand{\jbnote}[1]{}
\fi

\def\mean#1{\left< #1 \right>}

\permission{\copyright 2017 International World Wide Web Conference Committee \\ (IW3C2), published under Creative Commons CC BY 4.0 License.}
\conferenceinfo{WWW'17 Companion,}{April 3--7, 2017, Perth, Australia.}
\copyrightetc{ACM \the\acmcopyr}
\crdata{978-1-4503-4914-7/17/04. \\
http://dx.doi.org/10.1145/3038912.3038914 \\
\includegraphics{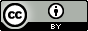}
}
 
\begin{document}

\title{I Would Not Plant Apple Trees If the World Will Be Wiped: Analyzing Hundreds of Millions of Behavioral Records of Players During an MMORPG Beta Test}

%
%
%
%
%

\numberofauthors{1}
\author{\alignauthor Ah Reum Kang$^\dagger$, Jeremy Blackburn$^\ddagger$, Haewoon Kwak$^\S$, Huy Kang Kim$^\P$\\
\affaddr{$^\dagger$University at Buffalo}\hspace{0.1cm} \affaddr{$^\ddagger$Telefonica Research} \\ \affaddr{$^\S$Qatar Computing Research Institute, Hamad Bin Khalifa University}\hspace{0.1cm} \affaddr{$^\P$Korea University}\\
\email{ahreumka@buffalo.edu, jeremyb@tid.es, haewoon@acm.org, cenda@korea.ac.kr}
}

\maketitle
\begin{abstract}
In this work, we use player behavior during the closed beta test of the MMORPG ArcheAge as a proxy for an extreme situation: at the end of the closed beta test, all user data is deleted, and thus, the outcome (or penalty) of players' in-game behaviors in the last few days loses its meaning.
We analyzed 270 million records of player behavior in the 4th closed beta test of ArcheAge. 
Our findings show that there are no apparent pandemic behavior changes, but some outliers were more likely to exhibit anti-social behavior (e.g., player killing).
We also found that contrary to the reassuring adage that ``Even if I knew the world would go to pieces tomorrow, I would still plant my apple tree,'' players abandoned character progression, showing a drastic decrease in quest completion, leveling, and ability changes at the end of the beta test.
\end{abstract}

%
%



%
%

%
%



\section{Introduction}
One problem that philosophers have struggled with over the centuries is how humans will behave in a disastrous ``end times'' scenario.  
For example, how does an individual behave if his/her behavior  will have no lasting outcomes or penalties?
Do we continue to follow the compass that has led us through life or do we abandon our morals, ideals, and social norms in the face of oblivion?
In this paper, we examine such a scenario through the lens of a massively multiplayer online role playing game (MMORPG).

In contrast to typical studies on running MMORPGs, our dataset is from the Closed Beta Test (CBT) of ArcheAge, developed and serviced by XLGames in Korea. 
The CBT is populated with a limited number of testers, and more importantly, at the end of the CBT the server is wiped: all characters are deleted, progression is lost, virtual property is deleted, etc.
The mapping principle~\cite{williams2010mapping} states that the behavior of players in online games is not very far from the behavior that humans exhibit in the real world.
Thus, while not a perfect mapping, we believe that the end of the CBT is a relatively good approximation of an ``end times'' scenario, and thus the present work is not only useful for the understanding of \emph{players'} behavior but can also begin to shed light on human behavior in general under such conditions.


From the ``living laboratory'' of the CBT, we first formulate the research problem. Our aim is to characterize the activity patterns of players over time with respect to a salient event.
A salient event is defined as an \emph{event} that takes place in \emph{time and space} which has \emph{impact} on \emph{social units} which can \emph{respond} to the event.
The closing of the CBT can be considered a salient event for the entire population of players. 

In this work, we investigate how player behavior changes during the course of the CBT.
We examine player behavior from two different levels, system-wide and individual-level, which have different granularity.
We do this to avoid an ecological fallacy~\cite{ess2001culture}, which is when statistical inferences about individuals are deduced from those about groups that they belong to.
Via a two-level analyses, we find no apparent pandemic (system-wide) behavior changes, although some outliers resorted to anti-social behavior, such as murder (player killing, or ``PK'').
That said, we surprisingly find that chat content exhibits a slightly \emph{positive} trend as the CBT draws to a close.
Overall, players increase social interaction with others: they exchange more in-game messages (mails) and create more parties to enjoy group-play or complete high-level quests. 

Additionally, we focus on whether individuals' behavioral changes are due to the CBT ending by comparing behavior to that of typical churners.
We find significant differences between players that voluntarily leave the game (churners) and those who stay until the end of the CBT.
In particular, we find that churners were more likely to exhibit anti-social behavior (e.g., PK). It seems that churners lose their their sense of responsibility and attachment to the game. In contrast, those who stay until the end might have some loyalty to the game and thus continue to behave within accepted social norms. 

Using network analysis, we focus on an associations between a wide range of individual player behaviors.
We examine patterns of in-game actions, looking for changes in the frequency as the CBT ends, find that contrary to the reassuring adage ``Even if I knew the world would go to pieces tomorrow, I would still plant my apple tree'' (i.e., I would still continue to better myself and the world), players abandoned character progression, showing a drastic decrease in quest completion, leveling, and ability changes. 
This finding itself is interesting and indicates why the quote resonates, and at the same time, it sheds light on game design implications for CBTs with respect to player reactions to the inevitable end of the beta test.

Our contributions are three-fold: 1)~we prove that analyses with different granularity, both individual-level and system-wide, are crucial to understand user behavior comprehensively; 2)~we propose a robust method to distinguish the effect of the end of the beta test from that of voluntary quit from the game by dealing with the typical churners separately; and 3)~to the best of our knowledge, we are the first to perform a large-scale quantitative characterization of behavior changes as the beta test of a game ends. This brings practical implications to game designers and theoretical implications to researchers who are interested in user behavior around a critical event.  


\section{Background}

\subsection{MMORPG as a Miniature of the World}

MMORPGs, such as World of Warcraft (WoW) and Ever Quest II, where a very large number of players interact with one another within a virtual world are one of the most popular forms of online gaming.
In MMORPGs, players choose a character and its role, race, and other traits, and live in the virtual world taking various actions.

ArcheAge is a medieval fantasy MMORPG serviced by XLGames\footnote{http://archeage.xlgames.com/en}.
Archeage has been released in Asia, Europe, and North America, and has over 2 million subscribers as of October 2014 \footnote{https://goo.gl/HHqCTx}. 

A design goal of ArcheAge is to offer players a playground to do whatever they want and find their own way to enjoy the game.
To this end, a wide range of actions are possible in ArcheAge compared to other MMORPGs.
Users can modify the environment by means of construction and cultivation much more extensively than in other comparable games. Users can enjoy player-vs-environment (PvE) combat and player-vs-player (PvP) combat, questing, group activities, chatting, housing and farming, crafting, trading, politics, voyaging and pirating, etc. 
All these actions are recorded in server-side game logs. Therefore, the logs are good assets to capture and observe more varied human behaviors than other games. 

\subsection{Closed Beta Test}

Online games development usually follows an ordered release process: Alpha Test, CBT, Open Beta Test (OBT), and then the official launch.
CBTs are private with a limited number of testers in order to find bugs, validate the market and fun of the game before official release, and improve the game through feedback.
A game can have multiple CBTs if necessary, and ArcheAge had five CBTs before launch.






\section{Related Work}

Castronova has put forth the theory that video games can serve as living laboratories allowing for novel social science research~\cite{castronova2006gamesasexperiments}.
The rules of the game world are not just explicitly known, but in fact completely controlled by the game developers.
This in turn allows us to study what amounts to real world behavior but still have some well defined controls.
In a nutshell, his thesis is that the scale, breadth, and depth of online games results in genuine social interactions, providing detailed, precise, and accurate traces at the society level.
Related to this, Williams~\cite{williams2010mapping} presents the mapping principle which posits that behavior in online video games ``maps'' to behavior in the real world.
I.e., that we can gain an understanding of real world behavior by examining behavior in online games.

Castronova et al.~\cite{castronova2009realasreal} further used traces of virtual goods transactions to measure whether there is a mapping principal for economic theories.
They quantitatively examine several economic indicators at large-scale based on the accurate, complete digital traces collected from an MMORPG.
Their findings show that virtual world denizens operate in the same way as in real world economies.
Like the work presented in this paper, having access to detailed data eliminates many of the concerns and limitations of traditional survey based studies.

Detailed game logs collected from the CBT of ArcheAge enable us to formulate research questions looking into players faced by an extreme situation. 
As previous efforts on online game beta tests have focused on things like market understanding~\cite{Gold01082012} or performance testing~\cite{jung2005venus}, we believe that the analogy between the server shutdown and an end times scenario presents a novel research challenge.



One well-known study around a critical event in online games was performed by Boman and Johansson, modeling a synthetic plague in WoW~\cite{magnus}.
The synthetic plague grew from what was originally conceived as a ``debuff'' intended to spread only from monsters to players.
A programming bug, however, resulted in the plague being able to spread from player to player.
As players constitute a synthetic society in the game, the game can be seen as an interactive executable model for studying disease spread (with the caveat that it is a very special kind of disease)~\cite{lofgren2007untapped}.
One interesting emerging behavior was that players would deliberately attempt to infect others by passing the debuff to their pets, dismissing them, and then re-summoning them in a populated area, causing the plague to spread.
Similarly, some industrious players set up sales of fraudulent cures.

Although this behavior is clearly anti-social, other behavior emer- ged to counteract it.
Some players acted as public health workers, healing the sick, while others even attempted to quarantine themselves and suffer a solitary death. This incident, however, is very limited in scope, which limits its implications.
In contrast, our work is applicable to any games that have beta tests, which are essential for game development. We, thus, believe that our work is easily generalizable with broad implications.  
Although ArcheAge has been studied previously and shown to be rich enough to capture complicated social dynamics~\cite{kang2013loyalty, kang2015twogames}, these previous works do not focus on the end games scenario.

\section{Dataset}

We acquired the anonymized full logs of the 4th CBT of ArcheAge directly from XLGames.
While we have the logs of the first two weeks of the 5th CBT, in this work, we focus on the 4th CBT because our aim is to examine user behavior as the CBT closes.
The logs were delivered as a 45 GB MySQL database and include essentially all actions that players take during the 4th CBT. For example, experience points gained, spells and abilities used, items purchased or crafted, etc.
The CBT took place between December 8th, 2011 to February 20th, 2012 (about 11 weeks) and is summarized in Table~\ref{tbl:log-files}.

\begin{table}[hbt]
\begin{center}
\small \frenchspacing
\begin{tabular}{cc}
\toprule
Period & 12/08/2011 $\sim$ 02/20/2012 \\
\midrule
\# of records & 275,274,108 \\
\# of players & 81,174  \\
\bottomrule
\end{tabular}
\end{center}
\caption{Summary of our dataset.}
\label{tbl:log-files}
\end{table}


We classified 75 different in-game actions into 11 categories: combat, party (grouping up with other players), instance dungeons (specially built dungeons with different content than the ``surface'' world), battle ground (a team deathmatch between players regardless of their race that takes about 15 minutes), siege warfare (battle between guilds over player owned castles; the largest scale collaborative play in the game), raid (a large party formed to defeat difficult boss enemies), expedition (ArcheAge's version of guilds), PvP, ``interaction doodad'' (players interacting with various objects in the world, e.g., harvesting a tree for wood), item production, and housing. This variety of in-game actions lets us study complex dynamics of a virtual world.

Finally, with access to such detailed data, we respect and protect privacy of game players.
All the data are anonymized. We have made no attempt to deanonymize it and do not have any information that can connect online and offline identity. 
Also, we note that our legal agreement with XLGames explicitly prevents us from receiving any information that can directly reveal players' real world identities.
We have further made a best attempt to avoid any analysis that would inadvertently reveal players' real world identity.

\section{Does a System-Level Pandemic E- xist?}

The first step in understanding how players deal with the end of the CBT is exploring their behavior evolved over time.
In this section, we examine aggregate player behavior in ArcheAge from the start of the CBT to the end.

\subsection{General In-Game Actions}

Like most MMORPGs, ArcheAge offers players a variety of activities to partake in.
In particular, ArcheAge has a sophisticated crafting system which allows players to produce items, a construction system allowing players to build houses, several types of PvE and PvP combat, as well as a variety of grouping and guild functionality.
However, not all of these actions are easily accessible at the start of the world.
For example, to build a house, players require some materials, such as a blueprint, wood, and ore.
To get a blueprint, players need to go to a specific region, called Mirage Island, and buy it from an NPC. 
For purchases, players need Gilda Stars, the main currency of ArcheAge, obtained by completing quests or trading packs.
To get wood, players must plant and harvest trees.
To get ore, players need to mine stone, metal, and gems from rocks. 
Overall, gathering these materials takes some play time.


\begin{figure}[t]
	\centering
	\includegraphics[width=1\columnwidth]{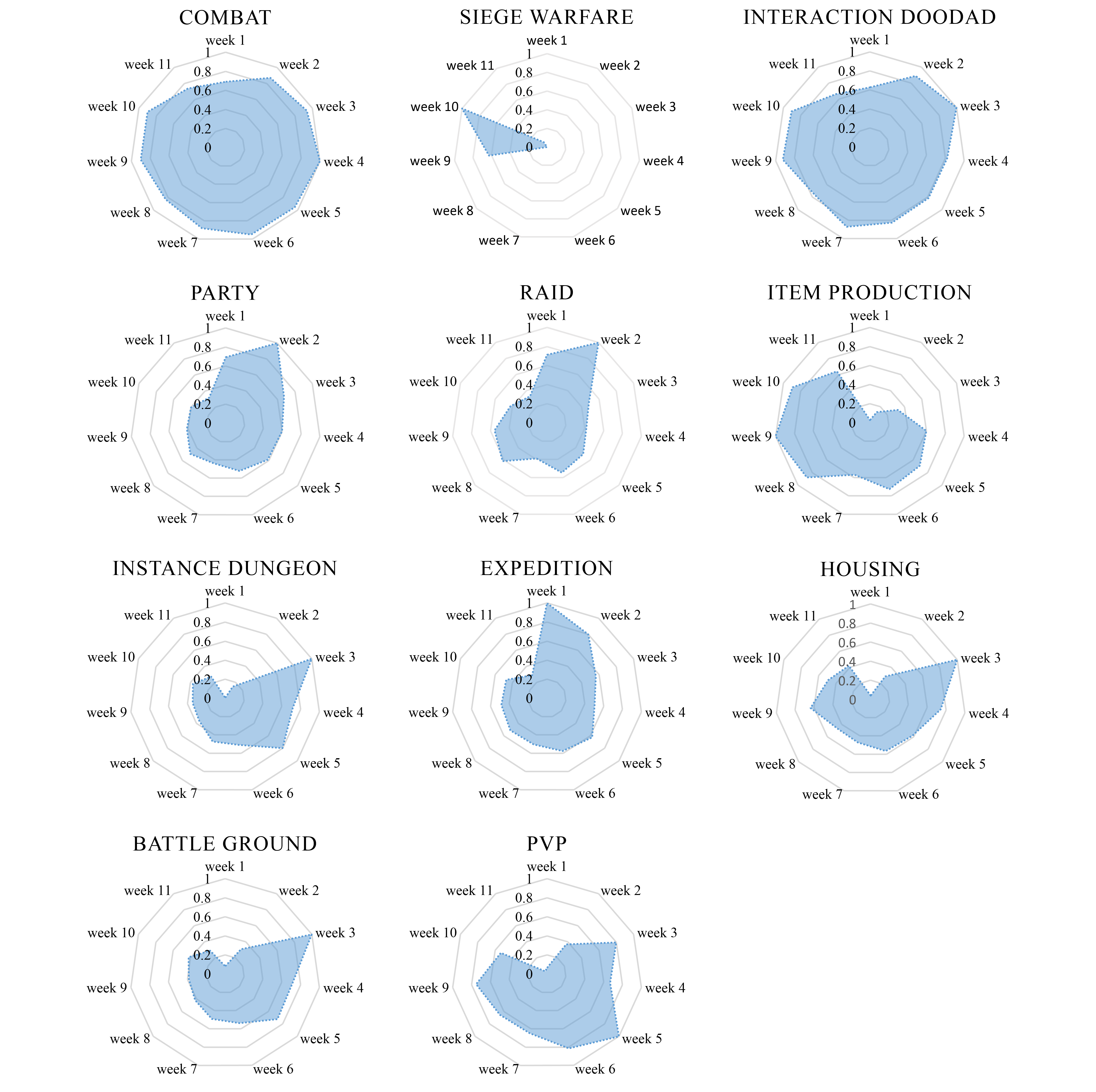}
    \vspace{-0.2cm}
	\caption{Radar plot of normalized action frequency per week. Weeks are arranged on the outer radius in clock-wise order. The further away the shaded region is from the center, the greater the frequency for the corresponding week.}
	\label{fig:action_change}
    \vspace{-0.2cm}
\end{figure}

Figure~\ref{fig:action_change} plots radar charts of the change in the frequency of actions per week, normalized by the number of users performing the action during the entire CBT.
Each week is demarcated by a tick on the outside radius (counter-clockwise ordering) and the further away the shaded region is from the center, the greater the number of actions.

From Figure~\ref{fig:action_change} we see that at week 1, expedition events were quite common.
Players are exploring and gathering information about the new world from the start.
Next, during week 2, party and raid events become popular.
This is likely due to players focusing on leveling up their characters with new found friends.
At week 3, we see many players taking part in instance dungeons, house building, and battle ground events.
Once players have established a foothold in the world, they begin to take part in more varied content for fun and currency. 
At week 4 and 5, we see a relative peak of item production and PvP activity. PvP requires rarer materials and in-game experience as it is generally considered more difficult than PvE combat.
Interestingly, after week 5, the frequency of PvP activities decline.
This shows that PvP tends not to be adopted as a regular activity by players. 
Siege warfare is observed at only week 9 and 10 because it was being tested during those periods only.
Overall, there are no extreme changes in in-game actions in terms of system-wide aggregated frequency over time, even though the CBT is ending.

\subsection{In-game Money Expenditure}

\begin{figure}[t]
	\centering
	\includegraphics[width=1\columnwidth]{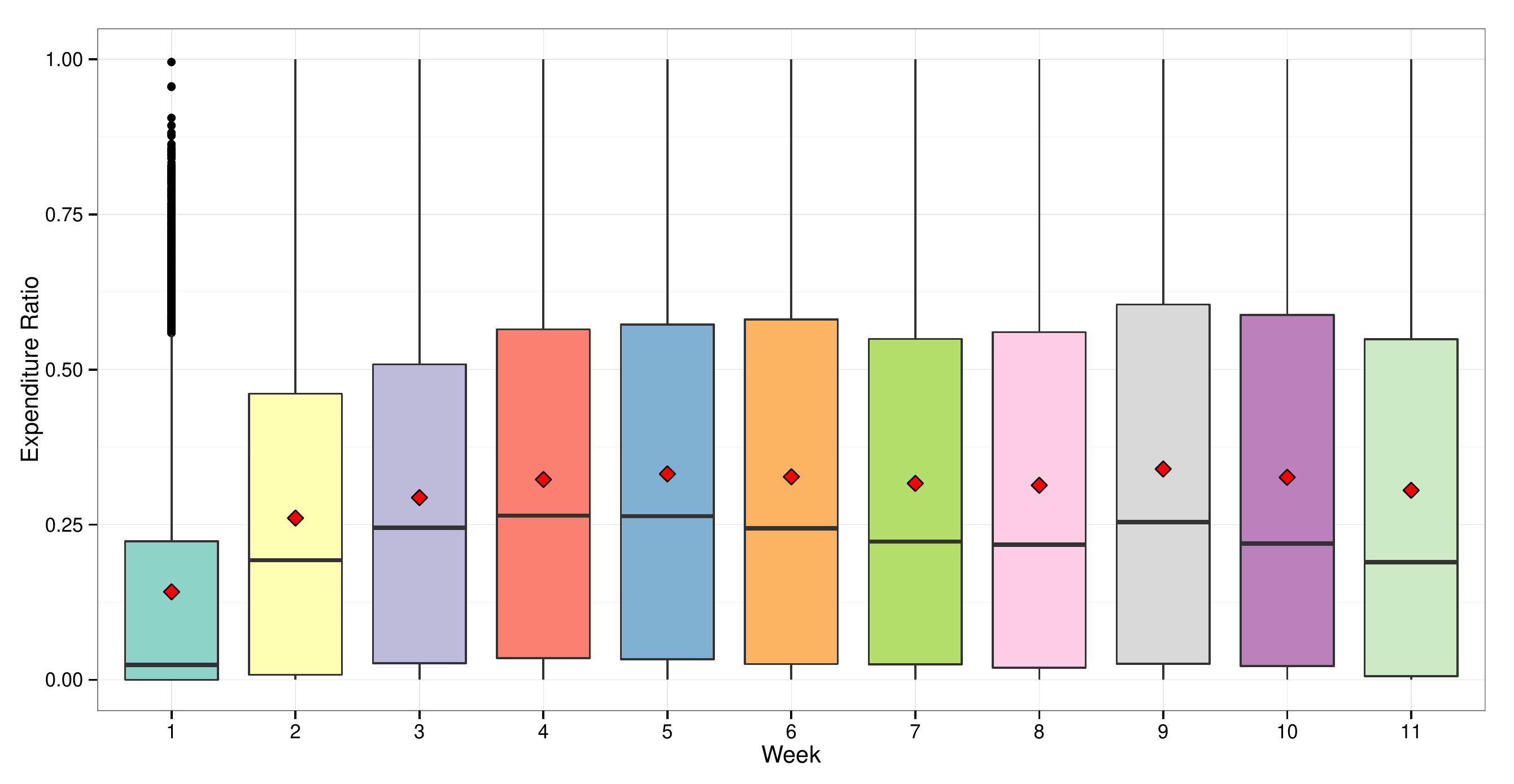}
    \vspace{-0.2cm}
	\caption{Distribution of expenditure ratio (money out flow divided by net worth) per week.}
	\label{fig:expenditure_rate_cdf}
\end{figure}

As mentioned earlier, ArcheAge has a strong virtual economy. Like other online games, ArcheAge players earn a currency, ``Gilda Stars,'' by killing monsters, complete quests, or trade items. Similarly, they spend money to buy housing, items, pets, or materials.

Intuitively, at the end of the CBT, there is no motivation to hold money in hand because it does not change to real money after the game ends. Thus, we hypothesize that players increase their rate of spending when the end approaches.
Figure~\ref{fig:expenditure_rate_cdf} plots the distribution of the expenditure ratio per week.
The expenditure ratio is computed as the amount of gold players spend divided by their net worth: if a player spends no gold they have an expenditure ratio of 0.0 but if they spend all the gold they have it is 1.0.

From Figure~\ref{fig:expenditure_rate_cdf} we see that in the earlier weeks players tend have a low expenditure ratio.
They are likely to save up resources to spend on higher-tier items like houses or boats.
Once they have built up their reserves, we see an increasing trend towards higher expenditure ratios.
At the 4th week, player expenditures become more stable, with perhaps a slight reduction as the end of the world approaches. This is opposite to our hypothesis. The potential explanation is that game players would not wait until the exact moment of the end of the CBT to spend their money to buy in-game items because it also requires some time to enjoy what they buy. Rather than the last week, we can see the higher peak at the 9th week, while the difference is marginal.

One interesting note is that a preliminary analysis on the economy of a different MMORPG (Aion\footnote{http://na.aiononline.com/en/}) found that players tended to have only around a 27\% expenditure ratio \cite{kang2013online}, whereas in Arche- Age the median is around 27\% but the average much higher. 

\subsection{Player-vs-Player Combat}

ArcheAge has a sophisticated PvP system to support a wide range of player actions.
Intuitively, we might expect that once it is apparent there are no consequences for actions, players might not feel the need to obey rules (e.g., ethics and social norms) and take part in anti-social behavior for fun.
ArcheAge's PvP system allows us to examine this: PvP between players of the same character race is classed as \textit{murder} with a variety of in-game consequences and penalties.

\begin{figure}[t]
	\centering
	\includegraphics[width=1\columnwidth]{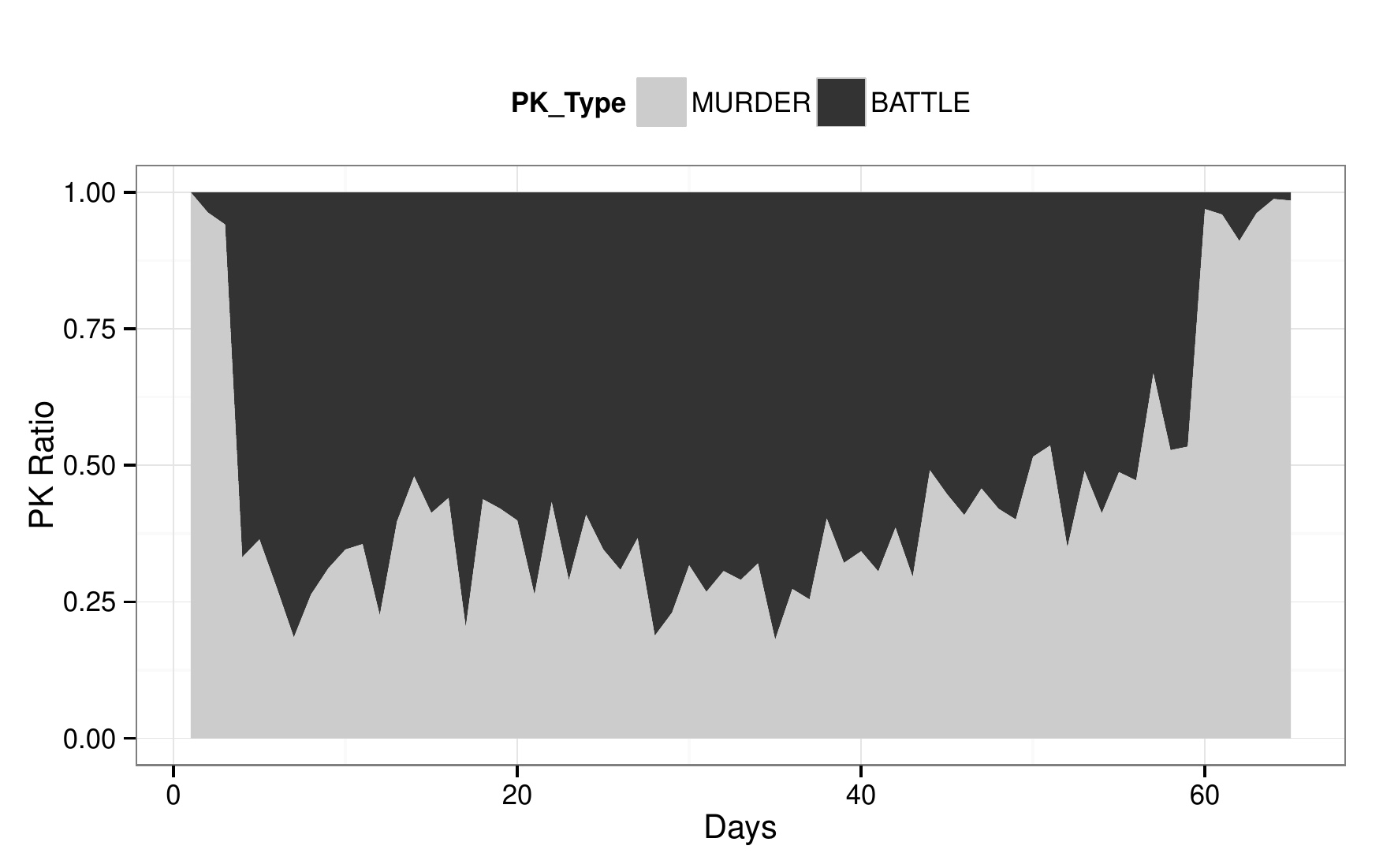}
    \vspace{-0.2cm}
	\caption{The relative frequency of murders over time.}
	\label{fig:the_rate_of_murder_events}
\end{figure}

Ultimately, we are concerned with whether or not players abandon whatever reservations they might have against murder.
Figure~\ref{fig:the_rate_of_murder_events} presents the relative frequency of regular PvP (battle kills allowed in the ArcheAge) to murders over time.
From the Figure, we see that murders are much more common at the beginning of the CBT, decreasing in proportion to regular PvP events until about the last third of the timeline where murders start becoming more prevalent again.
We suspect that the initial peak in murders is due to players trying out the PvP system as well as victimizing other new players who might not expect such early aggression.
The increasing trend at the end of the timeline is an indication that players might be reverting to more ``savage'' tendencies as well. As we expected, players are more likely to perform anti-social behavior when no penalty will be imposed.

Next we examine how pervasive such anti-social behavior was. We extracted all players that committed at least one murder in the last two weeks of the beta period.
We note that there were relatively few such murderers (334), and thus, the analysis does not represent pandemic behavior, but rather a closer look at outlier players that \emph{did} resort to violence at the end.

\begin{figure}[t]
	\centering
	\includegraphics[width=1\columnwidth]{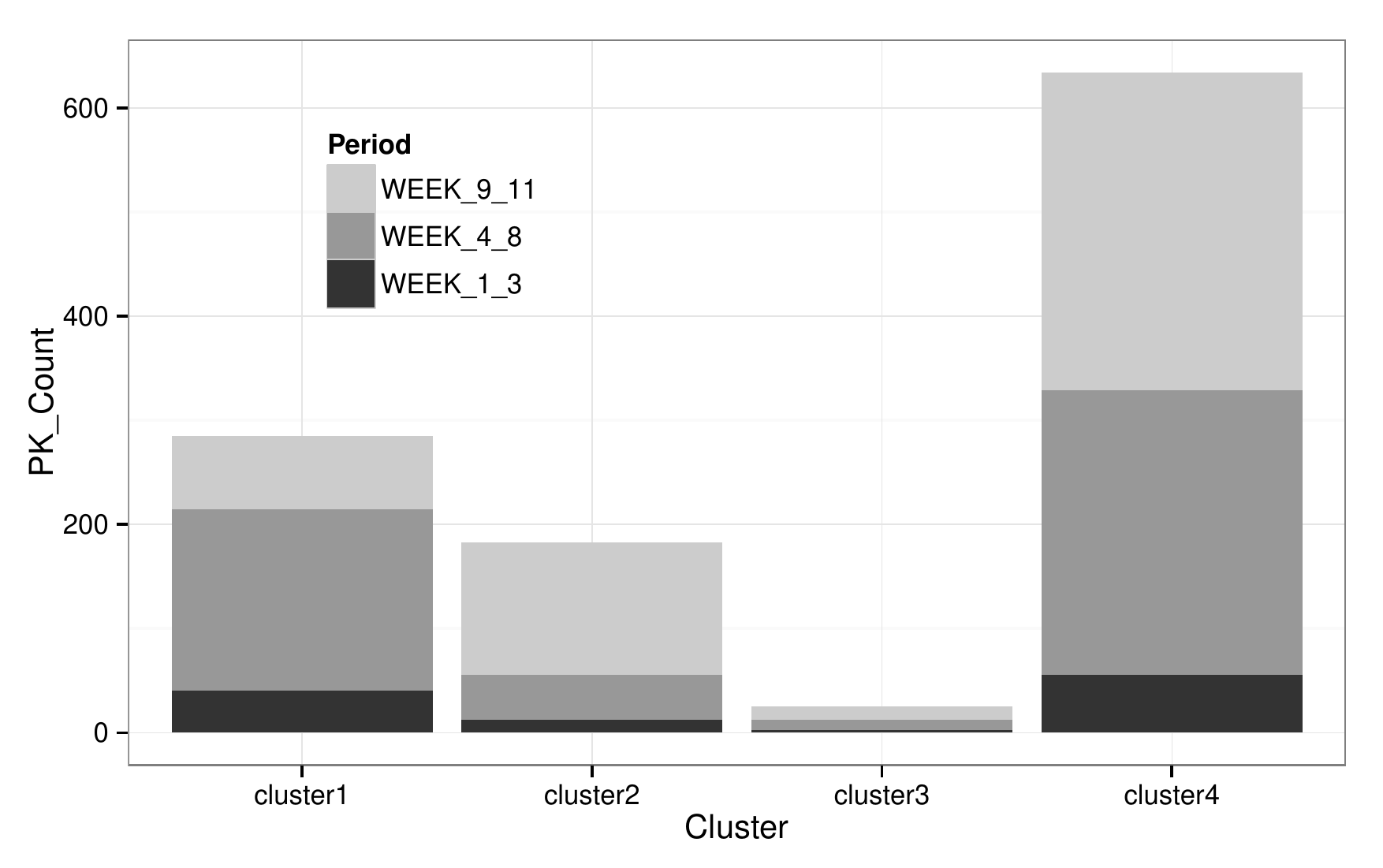}
    \vspace{-0.2cm}
	\caption{Frequency of murders per period committed by each cluster of murderers.}
	\label{fig:the_amount_of_murder}
    \vspace{-0.2cm}
\end{figure}

How do those 334 players behave during the whole beta test? Our interest here is whether they were ``normal'' players that turned into murderers only at the end of the CBT or whether they exhibited abnormal behavior throughout the CBT.
We cluster them based on their in-game activities, obtaining four clusters. We use $k$-means clustering algorithms and find the optimal $k$ by using an elbow method.

Figure~\ref{fig:the_amount_of_murder} plots a histogram of the murders performed per cluster divided into three intervals (early-, mid-, and end-times).
Interestingly, we see some differences in \emph{when} the different clusters performed their murders.
Even though the mid-times period encompasses more weeks than the end-times period, cluster 4 committed about the same number of murders in both periods and cluster 2 saw a dramatic \emph{increase} in murders.

There are two main takeaways from this finding: 1)~not all murderers are alike, but there do seem to be some archetypes they can be clustered under, and 2)~clearly there are some players who, although they did not quite go from pacifists to serial killers, did in fact show an increase in murderous tendencies as the end of the world drew near.

\subsection{Player Communication}

The second ``M'' in MMORPG stands for \emph{Multiplayer}, and thus the obvious draw to the genre is interacting with other players.
A big part of player interaction is social interaction via various forms of communication. ArcheAge supports a variety of real-time communication channels, many of which are tied to various action types.
For example, expeditions, parties, and factions all have their own chat channels.

These chat logs enable us to explore a few questions related to communication.
We have seen initial evidence that game play related behavior changes over time, but, does communication also change?


To answer this question, we examine emotions conveyed in chat messages via sentiment analysis. While many methods for sentiment analysis have been proposed, we use the simple, yet powerful, valence score~\cite{gonccalves2013comparing}.
A valence score is a measure of ``happiness'' and when applied to a corpus of text provides us with the sentiment expressed in the text.
As the ArcheAge CBT was populated almost entirely with native Korean speakers, we used a Korean language valence score dictionary compiled in~\cite{Dodds24022015}, which showed a slight, but statistically significant positive bias in human language across a variety of communication channels and languages. 
The valence scale ranged from unpleasant (1), to neutral (5) and pleasant (9).

\begin{figure}[t]
	\centering
	\includegraphics[width=1\columnwidth]{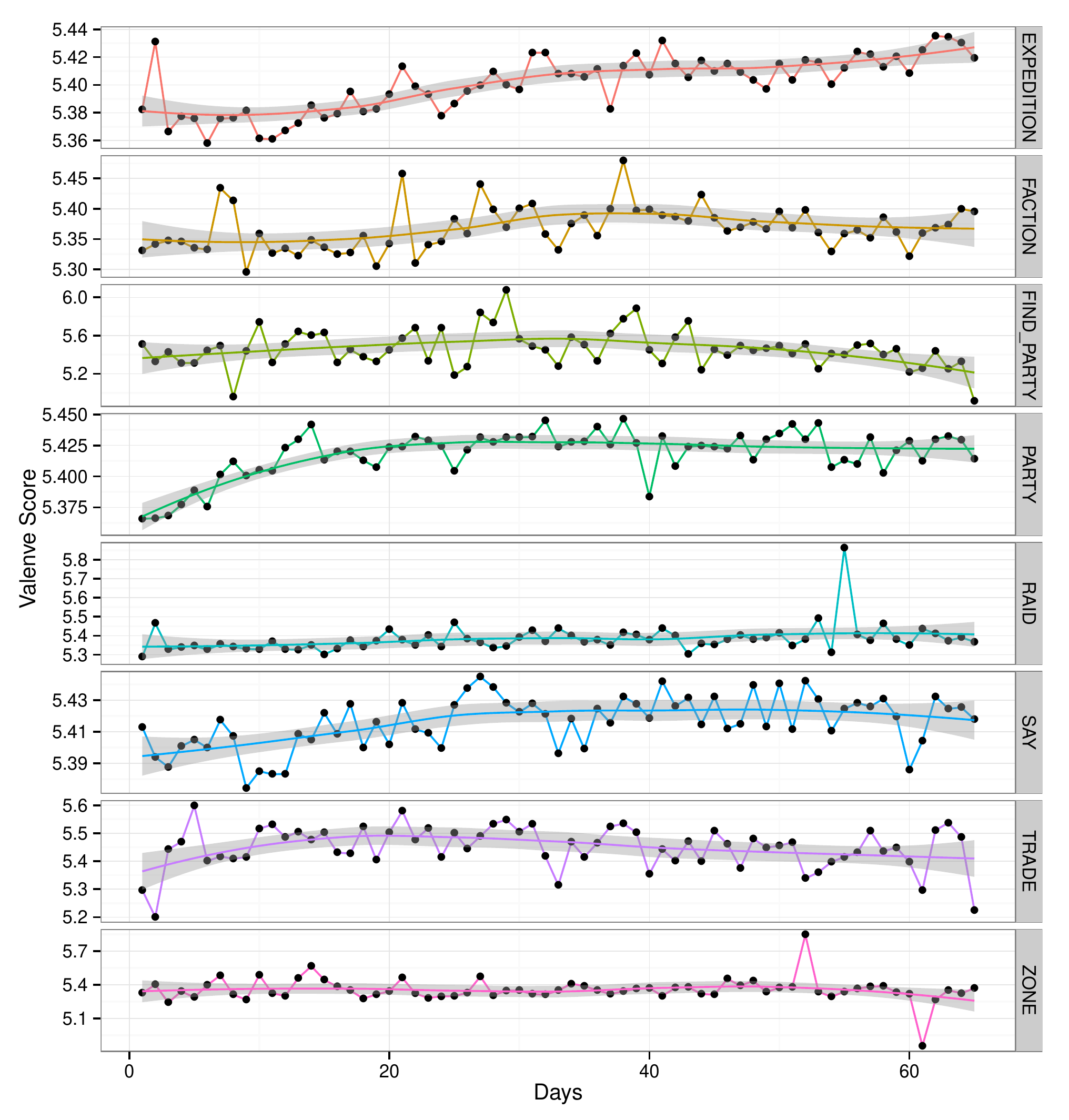}
    \vspace{-0.2cm}
	\caption{Valence scores over time by chat channel. Note differing y-scales.}
	\label{fig:chat_vscore}
    \vspace{-0.2cm}
\end{figure}

Figure~\ref{fig:chat_vscore} plots the valence scores over time for each public chat channel in ArcheAge.\footnote{We do not have the data of private channels.}
We plot the valence score for each day as well as a smoothed trend line.
We first note that as found in~\cite{Dodds24022015}, there is a slight positive bias across time; between 4.85 and 6.07 for the entirety of our dataset.
In particular, most of the valence scores hover right around 5.4, although there are some extreme outliers.
For example, the Raid channel saw a valence score of around 6.0 on one day, even though every other day was around 5.4.
We spoke to ArcheAge's developers and were unable to find an in-game reason (e.g., a special event put on by the game company) for this to happen, but we intend to look deeper into these outliers in the future.

More interestingly, we see that there are clearly different patterns expressed in the different chat channels over time.
For example, the Expedition channel has a clearly increasing ``happiness,'' while the Party channel has an increasing valence score for the first few weeks and then flattens out.
In general, the ``social'' channels (Expedition, Party, and Raid) see no decrease in valence as the end of the beta test approaches.
Although players do have somewhat changing sentiment, it is not sadness as the world reaches its end.



\section{Individual Player-level Behavioral Change}

So far we have examined aggregated dynamics from the system-wide view and found that global-scale pandemic behavior does not seem to emerge.  
In this section, we focus on each individual player to see whether there is a noticeable behavioral change that can be hidden when viewed in aggregate.

\subsection{The Measure of Irregularity of Player Behavior: Activity Peaks}

Our aim is to test whether an individual player shows behavioral change at the end of the CBT. To this end, we first define the irregularity of player behavior as the behavioral difference between the last day and the average day.
In other words, we generate time-series of the frequency of actions for each user and detect a positive (higher frequency than average) or negative peak (lower frequency than average) on the last day. 

\begin{figure*}[t]
    \begin{center}
    \includegraphics[width=2\columnwidth]{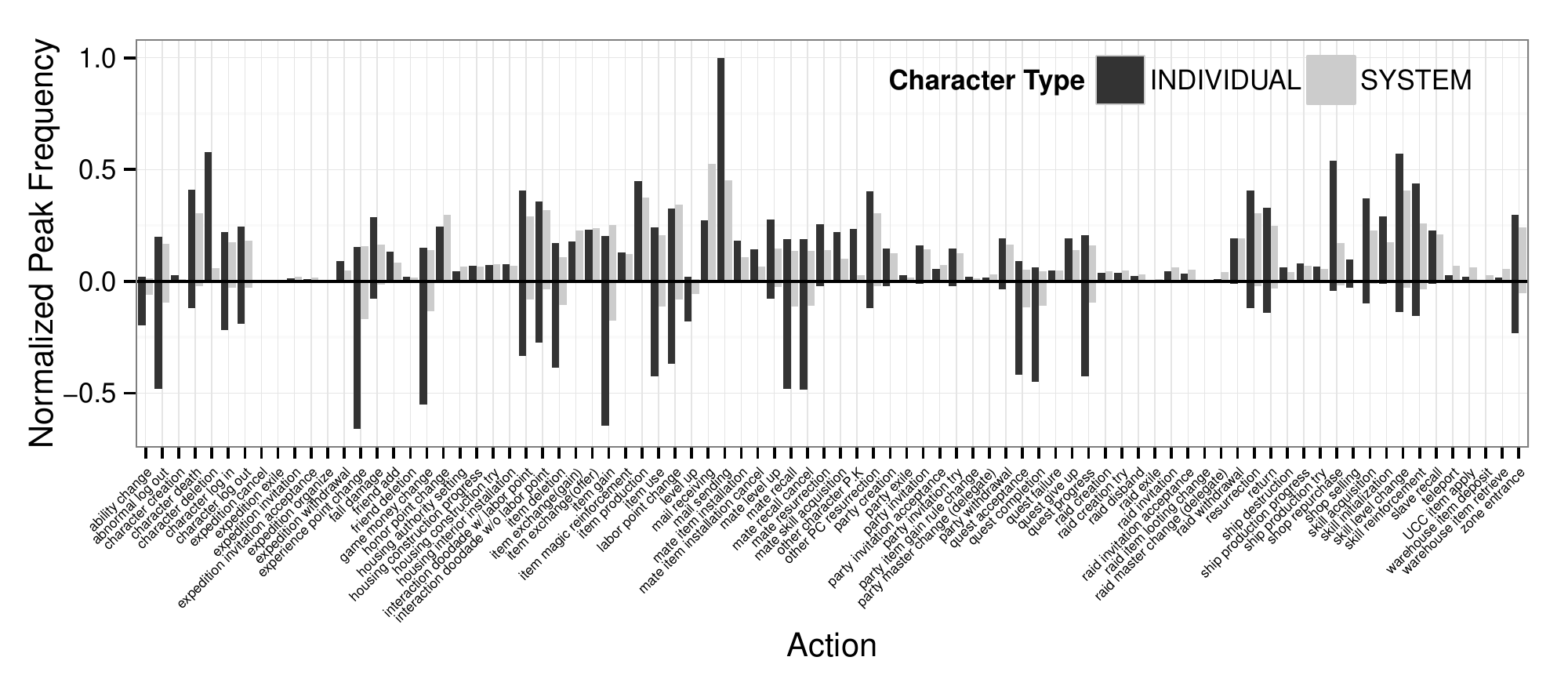}
    \caption{Normalized frequency of positive/negative peaks for each type of actions at the last day of the player's ``life.'' Negative peak frequency is shown on the negative y-axis. INDIVIDUAL players stopped playing \emph{before} the CBT ended, while SYSTEM players stayed until the end.}
    \label{fig:peaks-s-or-i}
    \vspace{-0.2cm}
    \end{center}
\end{figure*}

Although there are sophisticated approaches to detect peaks at arbitrary time point of a given streams like~\cite{kleinberg2003bursty}, 
what we need is much simpler.
We are interested in whether or not the irregular behavior is observed on the last day,
and thus, we take the straightforward approach by using the mean and the standard deviation of the time-series prior to the last login day.
More specifically, we define a \textit{positive peak} as the case where the frequency of a given action on the last login day of a player is greater than the sum of the mean frequency and two standard deviations of the action. Similarly, a \textit{negative peak} is defined as the case where the frequency of a specific action is less than the subtraction of two standard deviations from the mean frequency of the action on the last day.

More formally,
\[
peak(c, a_i) =
   \begin{cases}
       {+} & \text{if } |a_{i,\Omega}| > \mean{a_{i, 0..(\Omega - 1)}} + 2 \times \sigma_{i, 0..(\Omega - 1)}\\
       {-} & \text{if } |a_{i,\Omega}| < \mean{a_{i, 0..(\Omega - 1)}} + 2 \times \sigma_{i, 0..(\Omega - 1)} \\
   \end{cases}
\]
where $c$ is the player, $a_{i}$ is the action $i$, and $\Omega$ is the last day. While the height of the peak shows the strength of the irregularity, we first focus on the sign of the peak only.  

For the correct interpretation of peaks occurring at the last day, we compare them with peaks that can be observed from the players who leave the game earlier (i.e., voluntary churners). We denote the former with peaks for system-wide ($S$) reasons (i.e., the end of the beta test) and the latter with peaks for individual ($I$) reasons.  

The key question here is what are common and what are not common behaviors between $S$ and $I$ players because the circumstances that the players face are similar, but at the same time different. For example, players are not likely to care about the consequences of their actions on the last day because the next day does not come. In particular, penalties for bad actions, e.g., decrease in ``honor points'' or account suspension, are meaningless for both $S$ and $I$ players since they will not be playing another day anyways.

Nevertheless, there exist differences between the circumstances they face. It is whether \emph{others} are also supposed to leave the game \emph{at the same time and for the same reason}.
In other words, while $I$ users leave the game alone, by contrast, $S$ users leave the game together with all other players. From this difference we posit that system-wide impact could be observed from $S$ users but not from $I$ users. This means that some of the behavioral changes derived by shared emotions among the players might be observed from $S$ but not $I$.  

Also, there might be some differences between the attitude of the $S$ and $I$ users towards the game. Playing the game until the end of the beta test ($S$ users) shows players' devotion to the game. By contrast, leaving the game during the beta test ($I$ users) indicates that players lost interest. Losing interest in the game also connects to losing loyalty to the game and might lead to anti-social behavior. 

We filter out players that logged in less than 5 days (not necessarily consecutive) to exclude unstable one-time playing characters from our analysis. 
This filtering criterion leaves 22,945 players, 28.27\% of all players created during the 4th CBT.  
Among these relatively long-lived (again, $>$ 5 login days) players, we obtain 27,716 $peak(c,a_i)$ for 6,242 $I$ players and 15,430 $peak(c,a_i)$ for 3,150 $S$ players. 
Among them, 40.93\% of players show irregular behavior on their last login day.
In other words, four out of ten players behave irregularly on their last login day no matter whether leaving the game due to systemic or personal reasons.
This confirms our original intuition about the pervasiveness of unusual behavior on the last day.

Figure~\ref{fig:peaks-s-or-i} presents the number of players (either $I$ or $S$) that show positive and negative peaks on their last login day for each action normalized to $[0, 1]$ and $[-1,0]$, respectively. 
From the figure, we observe many differences between $I$ and $S$ players. 
Both $I$ and $S$ players show positive peaks for mail sending and receiving. In detail, $I$ players have more positive peaks for mail sending than receiving, but $S$ characters show more receiving than sending. 
Players who leave on purpose behave differently compared to players who leave due to the system end.

\begin{table*}[ht]
   \centering
{\footnotesize    
    \begin{tabular}{c c c c c}
            \toprule
             & \multicolumn{2}{c}{\textbf{SYSTEM}} &  \multicolumn{2}{c}{\textbf{INDIVIDUAL}} \\
            \textbf{Rank} & \textbf{Positive Peak} & \textbf{Negative Peak} & \textbf{Positive Peak} & \textbf{Negative Peak} \\
            \midrule
            1 & mail receive & level up & mail send & level up \\
            2 & mail send & ability change & mail receive & ability change \\
            3 & honor point change & expedition cancel & character deletion & quest completion \\
            4 & item production & quest completion & honor point change & quest acceptance \\
            5 & item exchange (offer) & quest acceptance & item production & experience point change \\
            6 & item exchange (gain) & expedition organize & PK & game money change \\
            7 & raid withdrawal & experience point change & mate item installation & expedition exile \\
            8 & mate item installation & expedition exile & item exchange (offer) & expedition cancel \\
            9 & party invitation & item deletion & mate item installation cancel & item gain \\
            10 & party creation & game money change & item exchange (gain) & mate recall cancel \\
            \bottomrule
    \end{tabular}
    }
\vspace{-0.2cm}
\caption{Top 10 positive and negative peaks for SYSTEM and INDIVIDUAL users.}
\label{tbl:top_peaks}
\vspace{-0.2cm}
\end{table*}

We then rank each action by its likelihood to show positive peaks and negative peaks. Considering the relatively small sample size of ArcheAge players compared to million-player scale MMORPGs, we use the lower bound of the Wilson score confidence interval for a Bernoulli parameter~\cite{wilson1927probable} instead of a simple ranking by the exact number of positive and negative peaks.
The method to compute the Wilson score is as follows:
\[
W=(\hat{p}+\frac{z^{2}_{\alpha/2}}{2n}-z^{2}_{\alpha/2}\sqrt{\frac{[\hat{p}(1-\hat{p})+z^{2}_{\alpha/2}/4n]}{n}})/(1+\frac{z^{2}_{\alpha/2}}{2})
\]
where $\hat{p}$ is the proportion of positive peaks to negative peaks, $n$ is the number of peaks, and $z^{2}_{\alpha/2}$ is the the (1-$\alpha/2$) quantile of the normal distribution. We use $z^{2}_{\alpha/2}$ = 1.96 for a 95\% confidence level.

Table~\ref{tbl:top_peaks} shows the top 10 actions with positive peaks (the highest $W$) and negative peaks (the lowest $W$) for $S$ players and $I$ players. From the top positive peaks of the $S$ players, we find an increase in social interaction, such as exchanging mails, party creation and invitation, and item exchange.  
The significant increment of item production shows that players willingly accept the risk of production failure at their last login days because the risk is not a risk any longer. 
Raid withdrawal indicates that raiding parties were broken up when the CBT ends. 

Among the top positive peaks of the $I$ users, we find two kinds of actions that are not observed from the $S$ users: character deletion and PK. As creating multiple characters is allowed for a single account, it is not essential to delete characters when leaving the game. Thus, it is reasonable to understand character deletion as intentional behavior for one's own reasons. We suppose that there is user intent to wipe out one's characters when they leave, perhaps related to growing privacy concerns online. 

PK observed as a positive peak is more interesting. ArcheAge churners lose their reservations and show online disinhibition just before they leave the game.
Even though PK is not explicitly banned by gameplay mechanics, it is fundamentally harmful to other players and involves risks such as penalty or item loss upon death. 
Several studies have examined the underlying conditions that catalyze toxic behavior~\cite{kwak2015exploring}, and leaving the game might very well be one more condition that can trigger toxic behavior. 
It is worth noting that by contrast, PK is not likely to appear (48th rank) for players who stayed until the end.

Most of the top negative peaks in Table~\ref{tbl:top_peaks} are common between the two sets of players. The easily recognizable trend is that players do not usually invest their time in making their characters better or stronger (e.g., level up, ability change, experience point change, quest accept, question complete, and game money change) once the end of the world approaches.
Players typically perform these actions because of their expected outcome in the future rather than an inherent enjoyment of the actions themselves.
Therefore, at the end of the beta test, they cease performing them.  
This is in contradiction to the adage, ``Even if I knew that tomorrow the world would go to pieces, I would still plant my apple tree.'' by Martin Luther
; people do not behave like that, and the counter-factual optimistic view might be why the adage resonated.

\subsection{Association of Peaks}

While Table~\ref{tbl:top_peaks} shows which action shows positive or negative peaks, players actually can do multiple actions. Then, how do such peaks associate with each other? 
To answer this question, we borrow some tools from network science and construct a \emph{network} of peaks and cluster the peaks into communities.

\begin{figure*}[hbt!]
  \begin{center}
\vspace{-10mm}  
  \includegraphics[scale=0.38]{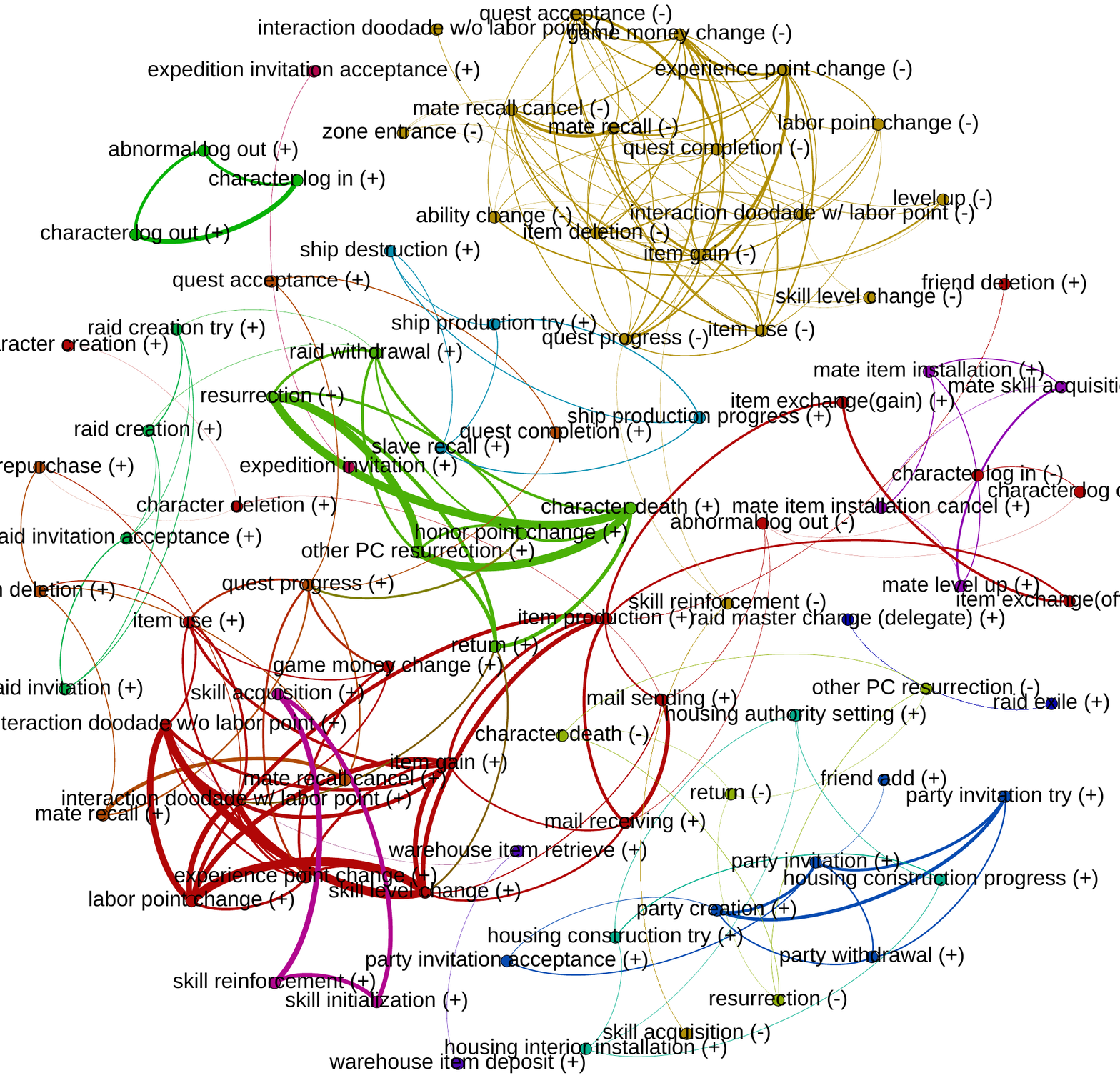}
  \includegraphics[scale=0.38]{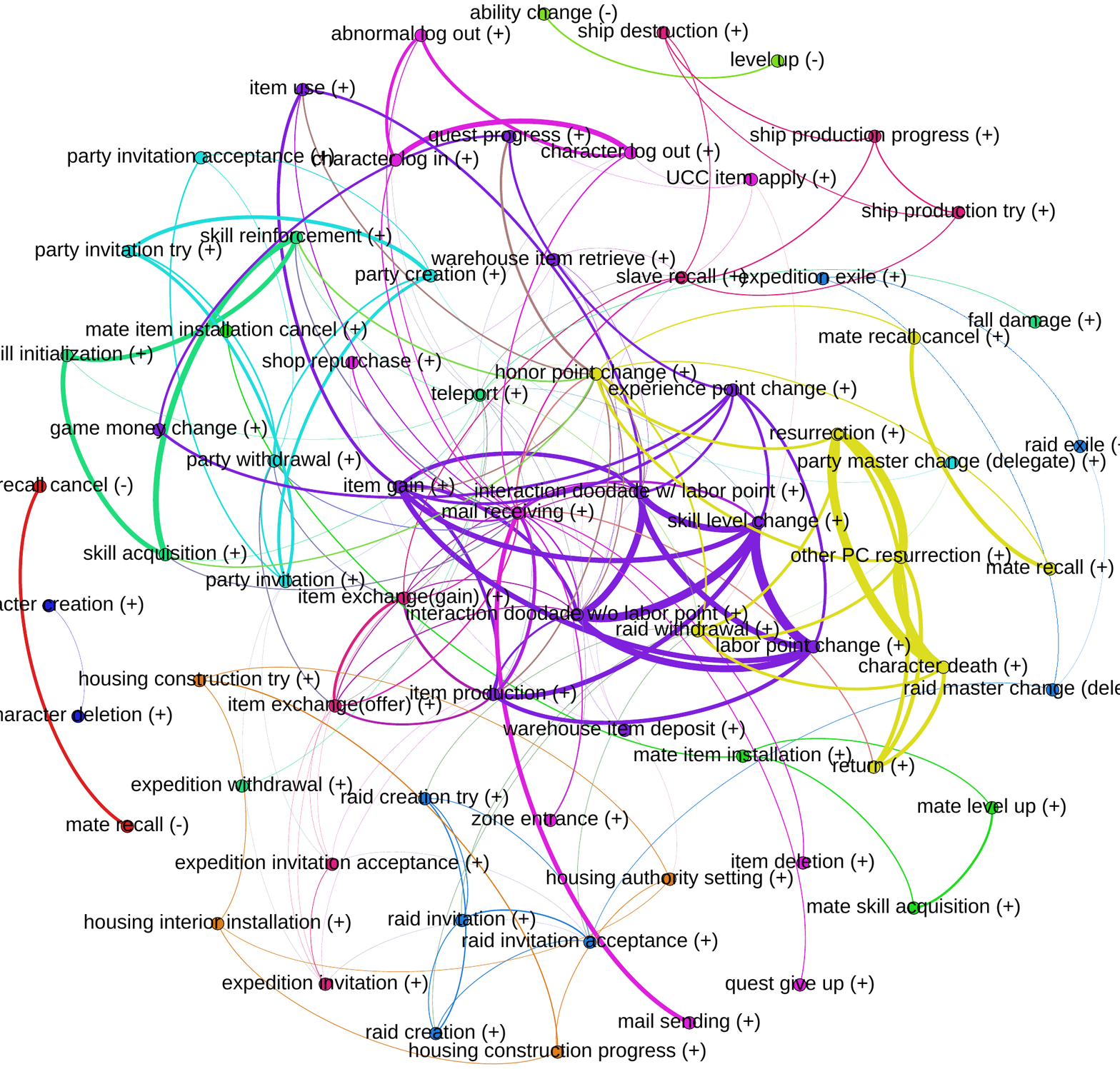}
\vspace{-10mm}  
  \caption{Backbone networks of peaks for users until the system ends (left) and users who personally leave before (right).}  
  \label{fig:backbone_networks}
  \end{center}
  \vspace{-0.2cm}
\end{figure*}

More formally, we build a graph $G(V, E)$ where each vertex is a pair of action and peak sign. 
I.e., $v_{a_i, sign} \in V$. 
For each vertex $v_i$, we define $c_{v_i}$ as the set of players that have the corresponding peak $v_i$. 
Next, we connect two vertices by an edge if and only if there are any players in common:
\[
e_{i,j} =
    \begin{cases}
        1 & \text{if } c_{v_i} \cap c_{v_j} \neq \emptyset \\
        0 & \text{otherwise}
    \end{cases}
\]

Naturally, we weigh each edge with the number of common players belonging to each vertex:
\[
w_{i, j} = |c_{v_i} \cap c_{v_j}|
\]

We note that vertices without edges (i.e., isolated vertices) are excluded from the graph. We also prune relatively unimportant edges based on their weights by backbone extraction~\cite{serrano2009extracting}. We set statistical significance level as 95\%. In pruning, the importance of a single edge can be measured differently depending on which of its vertices you measure from.
Thus, to better understand associations of peaks, we transform a single undirected edge into two reciprocal directed edges and prune them according to the statistical framework. In the remaining text, $G'$ denotes the pruned $G$.

We investigate two weighted directed networks, $G'_S$ and $G'_I$, built from $S$ users and $I$ users, respectively, which allows us to compare peak associations from the two group of users.
We find that $G'_I$ consists of fewer vertices than $G'_S$. It has 79 vertices and 464 edges, but $G'_S$ has 103 vertices and 498 edges. The difference mainly comes from negative peaks. Only four negative peaks (level up (-), ability change (-), mate recall (-), and mate recall cancel (-)) are included in $G'_I$, although 28 negative peaks are in $G'_S$. 

A large proportion of positive peaks in $G'_I$ means that players do more of those actions together at their last login day rather than doing them less, especially when they decide to leave the game.
More importantly, those positive peaks are well connected to each other like a clique. 

Intuitively, actions of a similar category are likely to synchronize. In other words, if one user does more leveling up, then the user is likely to do other activities related to making his character stronger. To support our intuition with data, we apply the Louvain method~\cite{blondel2008fast}, a widely-used modularity-based community detection algorithm, to identify densely connected local groups of peaks.

Figure~\ref{fig:backbone_networks} shows the communities derived from the peak networks of $G'_I$ and $G'_S$.
From this plot we make several observations.
We find 12 communities on $G'_S$ and 15 communities on $G'_I$. Low link density, defined as $|E|/\{|V|(|V|-1)\}$, of $G'_S$ (0.047) results in fewer communities, although the pruned network has more vertices than $G'_I$ (0.075).

Interestingly, we find no communities containing positive peaks and negative peaks at the same time in either network. In addition, this tendency, different signs of peaks never associated with each other, is confirmed by observing no direct edges between different signs of peaks in $G'_I$ and $G'_S$. 
I.e., players do not change their typical behavior in two different ways at the same time.

Also, communities from the two networks are not well overlapped, as we observed in Figure~\ref{fig:backbone_networks}. 
This finding is more evidence that $S$ players and $I$ players behave differently at their last login days. 
To quantitatively measure the similarity (or difference) between the two sets of communities, for each community in $G'_S$, we compute the Jaccard similarity coefficient with every community identified in $G'_I$. Then we pick the community with the maximum Jaccard coefficient as a \textit{corresponding} community for the specific community in $G'_S$. The average and the median of the computed Jaccard coefficient are 0.537 and 0.500, respectively. We note that in computing Jaccard similarity, we focus only on vertices (peaks) that are common in both networks because both networks have a different number of vertices.

\subsection{Changes in Player Communication}

\begin{figure}[t]
    \centering
    \includegraphics[width=0.95\columnwidth]{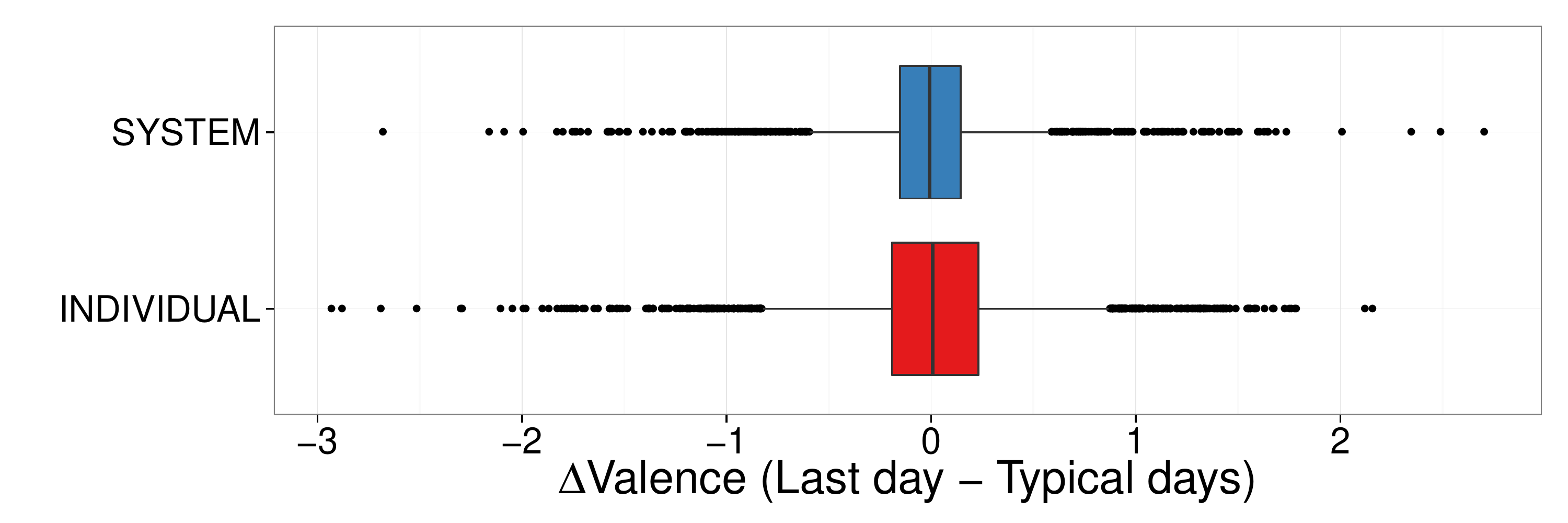}
    \vspace{-0.2cm}
    \caption{Change of valence score at the last day compared to typical days.}
    \label{fig:valence_score_change}
    \vspace{-0.2cm}
\end{figure}

In the system-wide view, we find that the sentiment of player chat slightly changes over time as in Figure~\ref{fig:chat_vscore}.
For our last experiment, we examine how an individual's chat behavior changes when their playing reaches an end. We again used valence score to capture sentiment.

Since the range of valence score that each player uses is different, comparing the average of the scores for all players during typical days and at the last day makes the behavioral change of each player average out. For example, the score of one user changes from 3 to 7 and that of the other user changes from 7 to 3. In this case, the average score does not change but stays at 5, even though each user shows substantial behavioral change. To address this issue, we track the change of every player. I.e., +4 and -4 for the two users in the above example.

For each player, we then subtract the average valence score of typical days from that of the last day and denote it by $\Delta$. Figure~\ref{fig:valence_score_change} shows a box-plot distribution of the $\Delta$ for the $S$ users and $I$ users. Although medians of the two groups are almost the same ($S$: $-0.01$, and $I$: $0.01$), we can see that distribution of $I$ users is much wider. In other words, users show more change in chat sentiment when they leave the game of their own volition than when the beta test ends.
Also, we have found a statistically significant effect of a group ($I$ or $S$) on the sentiment ($p < 0.05$). 

\begin{figure}[t]
    \centering
    \includegraphics[width=1\columnwidth]{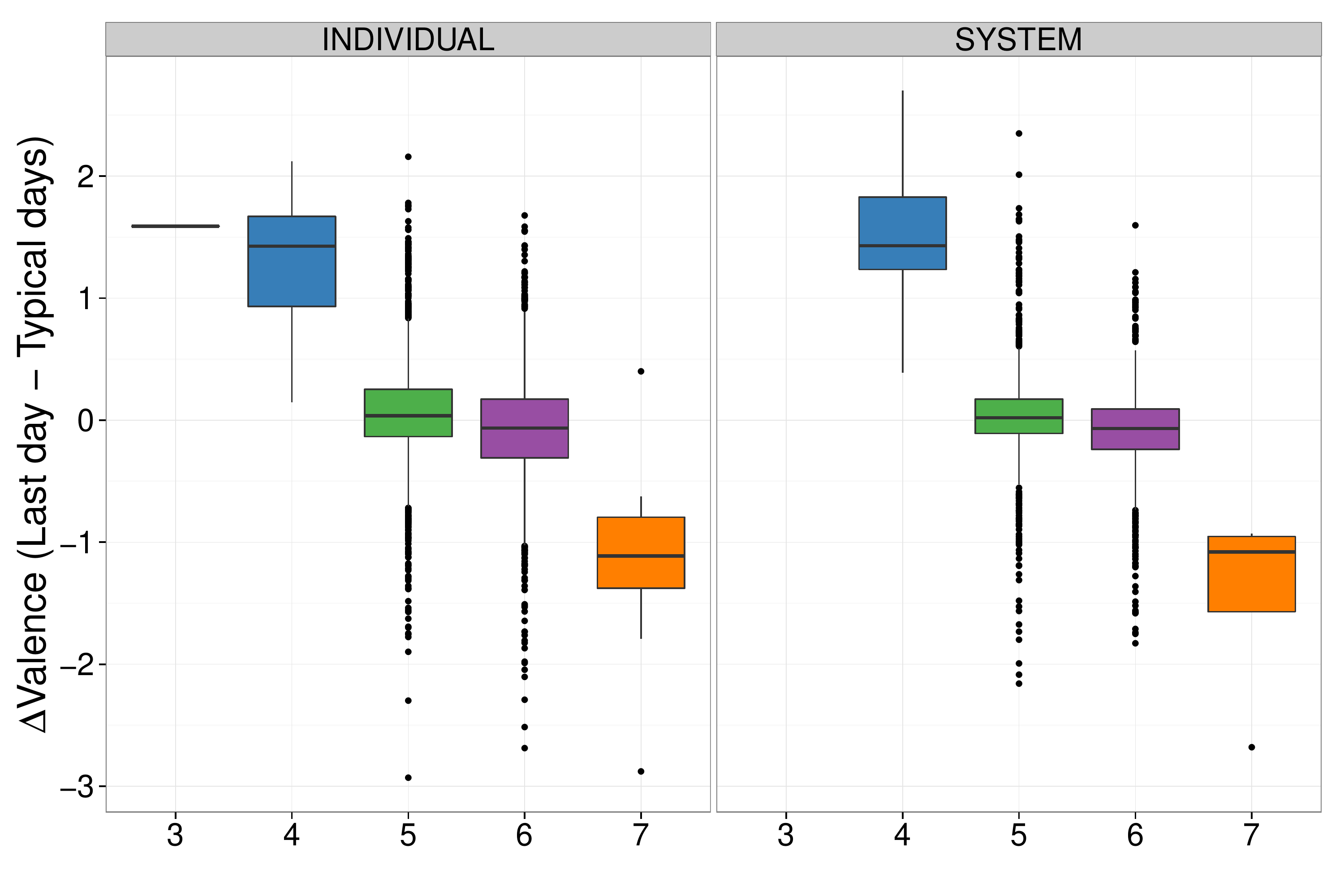}
    \vspace{-0.2cm}
    \caption{Change of valence score at the last day compared to typical days binned by the average valence score of typical days.}
    \label{fig:valence_score_change_agg}
    \vspace{-0.2cm}
\end{figure}

We break down the change of valence score based on players' typical valence score in Figure~\ref{fig:valence_score_change_agg}. 
We bin them by the rounded value of the average valence score for typical days and then show the distributions as a box-plot.  
By break down, an interesting pattern that is veiled in Figure~\ref{fig:valence_score_change_agg} emerges; $\Delta$ is positive when the sentiment of typical days is negative ($<$ 5) and negative when the sentiment of typical days is positive ($>$ 5). It means that the sentiment of the chats of an individual player at the last day is slightly positive no matter what the average sentiment of individual players' chat is.  

Also, combined with Figure~\ref{fig:valence_score_change}, we confirm that individual players' chat messages convey slightly positive attitude, with some exceptional outliers. This finding is consistent with previous studies on differences of chat messages between normal players and toxic players observed in North American and Western European players of a team competition game~\cite{blackburn2014stfu,kwak2014linguistic}. Therefore, this is additional evidence to support that 1)~sentiment analysis on chat messages is found to be appropriate for detecting outliers who show extreme behavior; and 2)~typical users' messages are slightly positive. The latter is also related to~\cite{Dodds24022015} which reports a universal positivity bias in human language. 

\section{Discussion and Conclusion}

In this work, we focused on understanding of user behavior during the beta test of an MMORPG.
We used detailed logs from the CBT of the MMORPG ArcheAge as a proxy for an extreme situation: at the end of the CBT, all user data is deleted, and thus the outcome of players' in-game behavior on the last days loses its meaning.
We examined this virtual petri dish in terms of how player behavior evolved over time and then focused on the particular behavioral changes that occurred right before the CBT ended.

Our findings show that there is no apparent pandemic behavior changes even when the CBT ends.
While we did find that some players resorted to anti-social behavior, such as murder, aggregate sentiment through chats shows pro-social trends.
When we focus on individual users' behavioral changes, we find significant differences between churners who voluntarily left the game before the end and players who stayed until the end.
In particular, we found that churners were more likely to exhibit anti-social behavior (PK).
Using network analysis, we found communities of related behavioral peaks. Interestingly, no communities contain positive and negative peaks simultaneously (i.e., no positive peak in a given behavior is correlated with a negative peak in a different behavior). Players do not change their typical behavior into two different ways at the same time.
Finally, when the CBT ends, we found that contrary to the reassuring adage, players abandoned character progression, showing a drastic increase in quest abandonment, leveling, and ability changes.

\subsection{Limitations}

Although we believe that our dataset represents about as close as we can get to an empirical end of the world scenario, there are several limitations to keep in mind.
First, ArcheAge is a video game and thus the true consequences of the CBT ending are entirely virtual: although plenty of in-game characters perish, no humans do.
Thus, it would be naive of us to claim a one-to-one mapping with real world behavior.
However, players do invest substantial time and energy into their characters, and it is quite common for virtual property to be worth real world money these days, so there are some real consequences.

Next, we examined data from the 4th CBT of ArcheAge, but there were three previous CBTs, one subsequent closed beta, and an open beta prior to the game officially launching.
It is quite possible that players become familiar with the end of the beta tests through the prior tests. However, the end of the particular instance of their virtual avatar affected their behavior.
That said, where relevant, our findings tend to be in agreement with those based on real world incidents, and thus we suspect this knowledge played little role.

\subsection{Implications}

Our study brings practical and theoretical implications to game industry and research communities.
Practically, our findings on irregular behavior of individual churners could be an alarm, or early-warning, of their leaving.
As addressing churners remains a consistent goal of game developers, our work can help inform the development of retainment strategies, such as offering incentives or new interactions to help them become attached to the virtual world. Also, what actions players increasingly or decreasingly perform when the end of the CBT comes provides guidance on how to run the CBT; some features should be tested earlier because players abandon them when the end of the CBT comes.

From the perspective of studying human behavior where
behavioral outcome does not have significant meaning, our findings that players do not invest their time for advancement and some outliers exhibit anti-social behavior can help design future studies. 

Also, we have provided additional empirical evidence in favor of the emergence of pro-social behavior.
Our findings that the sentiment of social grouping specific chat channels trend towards ``happier'' as the end times approach is a first indication of this pro-social behavior: existing social relationships are likely being strengthened.
Further, we saw that players that stayed until the end of the world exhibited peaks in the number of small temporary groupings: new social relationships are being formed.


\section{Acknowledgments}
This research was supported by Basic Science Research Program through the National Research Foundation of Korea (NRF) funded by the Ministry of Science, ICT \& Future Planning (2014R1A1A10 06228). 



%

\bibliographystyle{abbrv}
\bibliography{mybibfile}

\begin{thebibliography}{10}

\bibitem{blackburn2014stfu}
J.~Blackburn and H.~Kwak.
\newblock {STFU NOOB}!: predicting crowdsourced decisions on toxic behavior in
  online games.
\newblock In {\em Proceedings of the 23rd International Conference on World
  Wide Web}, pages 877--888, 2014.

\bibitem{blondel2008fast}
V.~D. Blondel, J.-L. Guillaume, R.~Lambiotte, and E.~Lefebvre.
\newblock Fast unfolding of communities in large networks.
\newblock {\em Journal of Statistical Mechanics: Theory and Experiment},
  2008(10):P10008, 2008.

\bibitem{castronova2006gamesasexperiments}
E.~Castronova.
\newblock {On the Research Value of Large Games: Natural Experiments in Norrath
  and Camelot}.
\newblock {\em Games and Culture}, 1(2):163--186, Apr. 2006.

\bibitem{castronova2009realasreal}
E.~Castronova, D.~Williams, {Cuihua Shen}, R.~Ratan, {Li Xiong}, {Yun Huang},
  and B.~Keegan.
\newblock {As Real as Real? Macroeconomic Behavior in a Large-scale Virtual
  World}.
\newblock {\em New Media {\&} Society}, 11(5):685--707, 2009.

\bibitem{Dodds24022015}
P.~S. Dodds, E.~M. Clark, S.~Desu, M.~R. Frank, A.~J. Reagan, J.~R. Williams,
  L.~Mitchell, K.~D. Harris, I.~M. Kloumann, J.~P. Bagrow, K.~Megerdoomian,
  M.~T. McMahon, B.~F. Tivnan, and C.~M. Danforth.
\newblock Human language reveals a universal positivity bias.
\newblock In {\em Proceedings of the National Academy of Sciences}, pages
  2389--2394, 2015.

\bibitem{ess2001culture}
C.~Ess.
\newblock Culture, technology, communication: Towards an intercultural global
  village.
\newblock {\em The Information Society}, 20(3):233--234, 2004.

\bibitem{Gold01082012}
S.~C. Gold and J.~Wolfe.
\newblock The validity and effectiveness of a business game beta test.
\newblock {\em Simulation \& Gaming}, 43(4):481--505, 2012.

\bibitem{gonccalves2013comparing}
P.~Gon{\c{c}}alves, M.~Ara{\'u}jo, F.~Benevenuto, and M.~Cha.
\newblock Comparing and combining sentiment analysis methods.
\newblock In {\em Proceedings of the First ACM Conference on Online Social
  Networks}, pages 27--38, 2013.

\bibitem{jung2005venus}
Y.~Jung, B.-H. Lim, K.-H. Sim, H.~Lee, I.~Park, J.~Chung, and J.~Lee.
\newblock Venus: The online game simulator using massively virtual clients.
\newblock {\em Systems Modeling and Simulation: Theory and Applications},
  3398:589--596, 2005.

\bibitem{kang2013loyalty}
A.~R. Kang, J.~Park, and H.~K. Kim.
\newblock Loyalty or profit? early evolutionary dynamics of online game groups.
\newblock In {\em Proceedings of the 12th Annual Workshop on Network and
  Systems Support for Games}, 2013.

\bibitem{kang2015twogames}
A.~R. Kang, J.~Park, J.~Lee, and H.~K. Kim.
\newblock Rise and fall of online game groups: Common findings on two different
  games.
\newblock In {\em Proceedings of the 6th Annual Workshop on Simplifying Complex
  Networks for Practitioners (collocated with WWW)}, pages 1079--1084, 2015.

\bibitem{kang2013online}
A.~R. Kang, J.~Woo, J.~Park, and H.~K. Kim.
\newblock Online game bot detection based on party-play log analysis.
\newblock {\em Computers \& Mathematics with Applications}, 65(9):1384--1395,
  2013.

\bibitem{kleinberg2003bursty}
J.~Kleinberg.
\newblock Bursty and hierarchical structure in streams.
\newblock {\em Data Mining and Knowledge Discovery}, 7(4):373--397, 2003.

\bibitem{kwak2014linguistic}
H.~Kwak and J.~Blackburn.
\newblock Linguistic analysis of toxic behavior in an online video game.
\newblock In {\em Proceedings of the First Workshop on Exploration on Games and
  Gamers (collocated with SocInfo)}, 2014.

\bibitem{kwak2015exploring}
H.~Kwak, J.~Blackburn, and S.~Han.
\newblock Exploring cyberbullying and other toxic behavior in team competition
  online games.
\newblock In {\em Proceedings of the 33rd Annual ACM Conference on Human
  Factors in Computing Systems}, pages 3739--3748, 2015.

\bibitem{lofgren2007untapped}
E.~T. Lofgren and N.~H. Fefferman.
\newblock The untapped potential of virtual game worlds to shed light on real
  world epidemics.
\newblock {\em The Lancet Infectious Diseases}, 7(9):625--629, 2007.

\bibitem{magnus}
S.~J.~J. Magnus~Boman.
\newblock Modeling epidemic spread in synthetic populations-virtual plagues in
  massively multiplayer online games.
\newblock In {\em Proceedings of Digital Games Research Association
  International Conference}, 2007.

\bibitem{serrano2009extracting}
M.~{\'A}. Serrano, M.~Bogun{\'a}, and A.~Vespignani.
\newblock Extracting the multiscale backbone of complex weighted networks.
\newblock In {\em Proceedings of the National Academy of Sciences}, volume 106,
  pages 6483--6488, 2009.

\vfill\eject
\bibitem{williams2010mapping}
D.~Williams.
\newblock The mapping principle, and a research framework for virtual worlds.
\newblock {\em Communication Theory}, 20(4):451--470, 2010.

\bibitem{wilson1927probable}
E.~B. Wilson.
\newblock Probable inference, the law of succession, and statistical inference.
\newblock {\em Journal of the American Statistical Association},
  22(158):209--212, 1927.

\end{thebibliography}

\end{document}